# Web Accessibility — A timely recognized challenge

**Jameel A. Qadri[1], M. Tariq Banday[2]**

[1]BC College of North West London, 470 Church Lane, Kingsbury, NW9 8UA, London,
e-mail: scorpiojameel@yahoo.com

[2]P. G. Department of Electronics and Instrumentation Technology, University of
Kashmir, Srinagar - 6, India, e-mail: sgrmtb@yahoo.com

**Abstract**

Web Accessibility for disabled people has posed a challenge to the civilized societies that claim to uphold the principles of equal opportunity and nondiscrimination. Certain concrete measures have been taken to narrow down the digital divide between normal and disabled users of Internet technology. The efforts have resulted in enactment of legislations and laws and mass awareness about the discriminatory nature of the accessibility issue, besides the efforts have resulted in the development of commensurate technological tools to develop and test the Web accessibility. World Wide Web consortium's (W3C) Web Accessibility Initiative (WAI) has framed a comprehensive document comprising of set of guidelines to make the Web sites accessible to the users with disabilities. This paper is about the issues and aspects surrounding Web Accessibility. The details and scope are kept limited to comply with the aim of the paper which is to create awareness and to provide basis for in-depth investigation.

## 1. Disability and Accessibility

### 1.1. Disability

'Disability' is any restriction or lack of ability to perform an activity in the manner or within the range considered normal for a human being (United Nations, 1983). According to UK's Disability Discrimination Act (DDA) 'a person has a disability if he has a physical or mental impairment which has a substantial and long-term adverse effect on his ability to carry out normal day-to-day activities. As per Americans with disability act (ADA), the term "disability" with respect to an individual stands to mean a physical or mental impairment that substantially limits one or more of the major life activities of such individual (ADA, 1990).

There are various types of disabilities that prevent the affected people to use the available technology without special purpose equipment generally known as assistive technology. WAI's Web Content Authoring Guidelines (WCAG) 1.0 has targeted following disabilities while drafting the standards for the web accessibility:

- **Visual:** People with visual disability include completely blind, low vision or color blind.
- **Auditory:** The people with auditory disabilities are hard of hearing or are completely deaf.
- **Motor Disabilities:** Disabilities like cerebral palsy, arthritis, Parkinson's disease or injury to spinal cord result in the decreased muscle movement or motor control making it difficult for the affected user to use the computer accessories like mouse and keyboard, the two main devices used to navigate through the websites.
- **Cognitive problems:** The cognitive problems include poor memory, problems in problem solving, linguistic problem involving difficulties in reading and verbal comprehension, and the lack of concentration.

## 1.2. Accessibility

'Accessibility', in context to the accessibility to the information and communication systems, specifically to the websites, refers to the extent by which the website, including the technology such as hypertext coding is barrier free to all users of information, thus providing enhancements that enable people with disabilities to move towards independence (Yates, 2005). Neal (2003) describes web accessibility as the ability for a person to understand and interact with the web content using hardware and/or software that renders the web content. Given the multi dimensional nature of the web which acts as source of information in the form of texts and multimedia, as an interacting technology and as a network interconnected with hyperlinks, World wide web consortium's (W3C) Web Accessibility Initiative (WAI) describes web accessibility to mean that people with disabilities can perceive, understand, navigate and interact with the web.

## 2. Web Accessibility Challenge

Since the advent of the Internet and its use as a medium for mass distribution of the information through World Wide Web (WWW), Internet has become inevitable for all and sundry. Internet has a role to play in everybody's life. According to Loo et al (2003) 'Internet use is expanding faster than any other communication technology in history and has the potential to significantly impact the major portion of the population in any society.' There is no known area where Internet cannot facilitate the processes involved. Whereas Internet is the technology that enables people to communicate with each other and transfer a huge mass of information, World Wide Web (WWW) is the actual product of this technology that has created an information revolution and has turned this world into a knowledge world. With Internet and WWW a new era has begun that has entirely changed the life style of people. However, Internet and Web while having facilitated the access to information on one hand but on the other hand has posed a danger of creating a digital divide in a society that has as its members disabled people as well. The challenge is how to make the information on Web reasonably available to every person irrespective of their limited capabilities to use the technology? Universal websites with accessible features are the likely solution.

Statistics show that the number of disabled people who are the potential users of the Internet and communication technology are not negligible, more in the light of the fact that the number of potential disabled users is likely to increase with the increased use of Internet. At the end of March 2006 the number of the working age population of registerable blind or partially sighted people in UK is estimated to be 80,000 (RNIB, 2006). According to the statistics provided by the Fred Hollows Foundation (FHF, 2007) as late as in April 2007, 14% of world's blind people live in Europe and America, the two developed regions in the world where Internet is used by 54% of the world population (IWS, 2008). Considering other forms of disability as well that prevent people to use the Internet and WWW, the number will be even more appalling.

With the proliferation of the Web and its growing use, the problem of exclusion of disabled people from using the Internet became grave. Efforts were started throughout world by governments, independent fora and other organizations to narrow down the digital divide between the normal users and the disabled people that eventually resulted in a number of enactment of legislations and laws in different countries, development of necessary technologies and web authoring tools, the development of standards to make the websites universally accessible and the accessibility evaluation tools to check conformity of websites to the set standards. United States passed ADA in 1990 and subsequently amended the section 508 of the Rehabilitation Act in 1998 addressing the issues of access to the sources of information by the disabled people. In UK Disability Discrimination Act was introduced in 1995 to protect the rights of the disabled people

## 3. Web Authoring and Assistive Technologies

Web accessibility directly depends on the tools that are used to develop a website called web authoring technologies and the technologies that interface the disabled users with machine are called assistive technologies.

### 3.1. Web Authoring Technology

Hyper Text Mark Up Language (HTML) is effectively used to develop a website that conforms to the set standards. In HTML the layout of the webpage is controlled with the help of formatting tags. The tags determine the visual presentation of the embedded elements. In the extensible HTML (XHTML) the presentational tags and attributes are removed from the individual elements and are collected separately on the HTML file or in a different file which is then included in the HTML page.

### 3.2. Assistive Technology

Assistive technologies include screen readers and magnifiers, closed captioning, alternative keyboards, and other special software and equipment that make information devices more accessible. The most discussed assistive technology in context of web accessibility is screen readers. Screen reader is software that presents the structured webpage as a linear document by reading the content of a computer screen aloud to the visually impaired user.

### 3.3. Evaluation Tools

Website evaluation is conducted on the websites to establish whether they conform to the web accessibility standards set by WCAG 1.0. The website accessibility evaluation tools check for the validation of the HTML code, accessibility, Screen Reader compatibility, Keyboard alternatives for images, animation and videos. The tool then displays a detailed report of the test indicating the conformance or otherwise to the set guidelines.

**Conclusion**

The Internet and web though being there now for more than a decade can still be considered as a recent phenomenon in the lives of people. The people who were already approaching old age when Web started found it difficult to use and understand the new technology because of its complexities. However, in present time the case is different and in future more people who will be in old age will be using Internet and Web as they are already familiar with it now being in their late middle ages. The inference is that much impairment that prevents the affected people to use the advanced technology like information technology appears and aggravates with the growing age, and in future the growing aged population who will be habitual users of Internet and Web will be on rise. So if accessibility is not sufficiently up, the agenda right now, but in future it will be just for the reason that it will affect far more people than it does today.

**Reference and Bibliography**